\documentclass[aps,twocolumn,superscriptaddress,showpacs]{revtex4}

\usepackage{graphicx}
\usepackage{epsfig}

\usepackage{epsfig,color,amssymb,amsbsy,bm}
\usepackage{amsmath}  
\usepackage{amsfonts} 
\usepackage{graphicx} 
\usepackage{stackengine}
\renewcommand{\bbox}{\boldsymbol }

\newcommand{\N}{Ne$_2$~}
\renewcommand{\H}{H$_2$~}

\newcommand{\ba}{\begin{eqnarray}}
\newcommand{\ea}{\end{eqnarray}}
\newcommand{\br}{\begin{eqnarray*}}
\newcommand{\er}{\end{eqnarray*}}
\newcommand{\be}{\begin{equation}}
\newcommand{\ee}{\end{equation}}
\newcommand{\nn}{\nonumber}
\newcommand{\ra}{\rangle}
\newcommand{\la}{\langle}
\newcommand{\bs}{\bigskip}
\newcommand{\bp}{\begin{minipage}}
\newcommand{\ep}{\end{minipage}}
\newcommand{\bt}{\begin{tabular}}
\newcommand{\et}{\end{tabular}}

\newcommand{\eref}[1] {(\ref{#1})}
\newcommand{\Eref}[1] {Eq.~(\ref{#1})}
\newcommand{\Fref}[1] {Fig. \ref{#1}}
\newcommand{\w}{{\omega}}

\newcommand{\Wcm}[2]{
$\rm {#1}\times10^{{#2}}~W/cm^2$}
  
\renewcommand{\k}{{\bbox k}}
\newcommand{\hs}{\hspace*}
\newcommand{\vs}{\vspace*}
  \newcommand{\A}{{\bbox A}}
  \newcommand{\E}{{\bbox E}}
  
\renewcommand{\r}{{\bbox r}}
\newcommand{\R}{{\bbox R}}

\begin{document}

\title{Numerical attoclock  on atomic and molecular hydrogen}

\author{Vladislav V. Serov}

\affiliation{ Department of Theoretical Physics, Saratov State
  University, Saratov 410012, Russia}

\author{Alexander W. Bray} 
\author{Anatoli S. Kheifets}
\email{A.Kheifets@anu.edu.au}
\affiliation{Research School of Physical Sciences,
The Australian National University,
Canberra ACT 0200, Australia}

\date{\today}
\begin{abstract}
Numerical attoclock is a theoretical model of attosecond angular
streaking driven by a very short, nearly a single oscillation,
circularly polarized laser pulse. The reading of such an attoclock is
readily obtained from a numerical solution of the time-dependent
Schr\"odinger equation as well as a semi-classical trajectory
simulation. By making comparison of the two approaches, we highlight
the essential physics behind the attoclock measurements. In
addition, we analyze the predictions the Keldysh-Rutherford model of
the attoclock [Phys.\ Rev.\ Lett.\ {\bf 121}, 123201 (2018)]. In
molecular hydrogen, we highlight a strong dependence of the width of
the attoclock angular peak on the molecular orientation and attribute
it to the two-center electron interference. This effect is further
exemplified in the weakly bound neon dimer.

\bs

\end{abstract}



\maketitle


The experimental technique of attosecond angular streaking (attoclock)
is based on measuring an offset angle of the peak photoelectron
momentum distribution (PMD) in the polarization plane of a
close-to-circularly polarized laser pulse. The attoclock attempts to
relate this offset angle with the time the tunneling electron spends
under the barrier (tunneling time)
\cite{Eckle2008,P.Eckle12052008,Pfeiffer2012,Landsman:14}. As the
tunneling is an exponentially suppressed process, it occurs
predominantly at the peak of the driving laser pulse. At this
instant, the  electric field is aligned with the major axis of the
polarization ellipse.  The photoelectron emerges from the tunnel with
the zero velocity and its canonical momentum captures the vector
potential of the laser field at the time of exit. This momentum is
carried to the detector and its angular displacement relative to the
minor polarization axis is converted to the tunneling time $\tau =
\theta_A/\w$, where $\w$ is the angular frequency of the driving field.
A similar attoclock reading $\theta_A$ can be obtained from numerical
simulations with very short, nearly single oscillation, circularly
polarized pulses. The utility of such a ``numerical attoclock'' is
that it allows for treatment by various simplified, but more
physically transparent, techniques such as an analytic $R$-matrix
theory \cite{Torlina2015}, a classical back-propagation analysis
\cite{PhysRevLett.117.023002}, classical-trajectory Monte Carlo
simulations \cite{0953-4075-50-5-055602} and a classical Rutherford
scattering model \cite{PhysRevLett.121.123201}. By making comparison
with these models, numerical attoclock experiments firmly point to a
vanishing tunneling time
\cite{Torlina2015,PhysRevLett.117.023002,PhysRevLett.121.123201}.
Similar conclusion was also reached in recent theoretical
\cite{PhysRevA.97.031402,Ivanov2018} and experiemntal
\cite{Sainadh2019} works.  The debate of the finite tunneling time is
still open. Some authors continue to advocate a finite tunneling
time \cite{Landsman20151,PhysRevLett.116.233603,PhysRevLett.119.023201} while others suggest
that the whole concept is ill defined and no meaningful definition of
the tunneling time can be given \cite{Sokolovski2018}.

Irrespective of  the answer to the  tunneling time conundrum, the
principle of attoclock remains appealing and finds its application to
more complex targets. In particular, there have been preliminary
reports of attosecond angular streaking measurements on molecular
hydrogen \cite{Satya2018,Wei2019}. Coincident detection of
photoelectrons and molecular fragments in dissociative ionization of
\H allows to conduct attoclock measurements on aligned molecules and
to explore the effect of molecular orientation.

To highlight the essential physics behind the attoclock measurements
on \H and to contrast them with analogous measurements on the hydrogen
atom, we employ the same principle of the numerical attoclock.  We solve
the time-dependent Schr\"odinger equation (TDSE) driven by a short,
nearly single-cycle, circularly polarized laser pulse. In parallel, we
simulate the numerical attoclock within the strong field approximation (SFA) by
performing the steepest descent integration using the saddle point
method (SPM) \cite{PhysRevLett.89.153001,0953-4075-39-14-R01}. With
such a short driving pulse, this analysis is streamlined and the whole
PMD can be obtained from the contribution of just a single strongly
dominant saddle point.

We find the attoclock offset angles of the H atom and the \H molecule
to be rather similar, the latter only weakly dependent on the
molecular axis orientation relative to the polarization plane. We
interpret these results within the so-called Keldysh-Rutherford (KR)
model \cite{PhysRevLett.121.123201} in which the photoelectron
undergoes elastic scattering on the Coulomb potential of the residual
ion. The point of the closest approach in this model is equated with
the Keldysh tunnel width $b=I_p/E_0$ expressed via the ionization
potential $I_p$ and the peak electric field $E_0$. At the laser pulse
intensities under consideration $E_0\ll1$~a.u. Thus the tunnel width
is significantly larger than the inter-atomic distance $b\gg
R$. Hence, the molecular orientation is of little effect. In contrast,
the angular width of the photoelectron peak, both in and out of the
polarization plane, is markedly different for H and \H\!\!\!. More
importantly, this width depends strongly on the molecular axis
orientation and bears a clear signature of the two-center electron
interference. Such an interference has been predicted theoretically
for single-photon molecular ionization
\cite{PhysRev.150.30,KapMar69}. It had been observed in various
molecular ionization processes and, more recently, in strong field
phenomena including high order harmonic generation and above threshold
ionization
\cite{PhysRevLett.85.2280,PhysRevLett.95.153902,PhysRevLett.95.093002,
  PhysRevLett.100.073902,PhysRevLett.101.053901,Augstein2012}.  This
effect is further exemplified in  weakly bound noble gas clusters
where it depends sensitively on the symmetry of the dissociative ionic
state \cite{Kunitski2019}. We confirm it here by conducting the SPM
calculations on the \N dimer.

The rest of the paper is organized as follows. In Sec.~\ref{Methods}
we describe our quantum-mechanical (\ref{Quantum}) and semi-classical
(\ref{Classical}) techniques. In Sec.~\ref{Results} we present the
results of our simulations on atomic (\ref{Atomic}) and molecular
(\ref{Molecular}) hydrogen. In Sec.~\ref{Atomic} we also compare the results
of numerical \cite{PhysRevLett.121.123201} and real \cite{Sainadh2019}
attoclock experiments and highlight the range of validity of the KR
model.  We also make a comparison of the TDSE and SFA-SPM predictions
for numerical attoclock. This comparison exemplifies the role of the
Coulomb field of the residual ion.  In Sec.~\ref{Molecular} we compare
the readings of the atomic and molecular attoclocks and attribute
their difference to the two-center Young-type interference. This
interference is manifested much stronger in a weakly bound and very
extended neon dimer \N which we consider in
Sec.~\ref{Dimer}. 
%

\section{Methods and techniques}
\label{Methods}

\subsection{Quantum mechanical simulations}
\label{Quantum}

We solve numerically the  TDSE 
\begin{equation}
\label{TDSE}
i {\partial \Psi(\r) / \partial t}=
\left[\hat H_{\rm target} + \hat H_{\rm int}(t)\right]
\Psi({\bbox r}) \ ,
\end{equation}
where $\hat H_{\rm target}$ describes a one-electron target in the
absence of the applied field and the interaction Hamiltonian is
written in the velocity gauge
\be
\label{gauge}
\hat H_{\rm int}(t) =
 {\bm A}(t)\cdot \hat{\bm p} \ \ , \ \ 
\bm{E}(t)=-\partial \bm{A}/\partial t \ .
\ee
Here the vector potential of the driving pulse is 
\be
{\bbox A}(t) = 
{A_0\over \sqrt{\epsilon^2+1}}
 \cos^4 (\w t/2N +\phi)
\left[
\begin{array}{r}
\cos(\w t)\\
\epsilon \sin(\w t)
\end{array}
\right]
\label{pulse}
\ee
with the ellipticity parameter $\epsilon$, the angular frequency $\w$
and the carrier envelope phase (CEP) $\phi$. The
pulse length is parametrized with the number of optical
cycles $N$ and ${\bbox A}$ vanishes for $|t| \ge N\pi/\w$. Note that
for $\phi=0$, the center of the pulse corresponds to $t = 0$.
The (peak) field intensity is given by $I=2(\omega A_0)^2$ and the
frequency $\omega$ is taken to correspond to $800$ nm radiation.  The
single-oscillation pulse employed in the present work corresponds to
$\epsilon=1$, $N=2$ and $\phi=0$. 
%
%
In contrast, multi-cycle pulses with $\epsilon=0.88$, $N=5$ and
various $\phi$ were employed in numerical simulations of the real
experiment \cite{Sainadh2019}.
%
%
It appears that the effect of the driving pulse shape
on the attoclock reading $\theta_A$ is very significant as will be
highlighted in Sec.~\ref{Atomic}.

In the case of atomic hydrogen, TDSE \eref{TDSE} was solved by two
different numerical techniques: the iSURF method
\cite{0953-4075-49-24-245001} and the split-operator method
\cite{PhysRevA.84.062701}. The iSURF method was used previously both
for numerical attoclock \cite{PhysRevLett.121.123201} and real
experiment \cite{Sainadh2019} simulations. A similar set of numerical
parameters was employed as in the previous works. The two TDSE codes
\cite{0953-4075-49-24-245001,PhysRevA.84.062701} were benchmarked
against each other and the numerical attoclock results were found to
be identical.  The split-operator method \cite{PhysRevA.84.062701} was
also adapted for molecular hydrogen. The \H molecule at the
equilibrium internuclear distance $R=1.4$~a.u. was treated within the
single active electron approximation on the frozen-core Hartree-Fock
(HF) basis \cite{PhysRevA.84.062701}.  The radial integration was
conducted by the finite-element method on the Gauss-Lobatto
quadratures. The radial grid parameters were chosen as follows: the
finite element order $N_{\rm ORD} = 4$, the number of finite elements
$N_{\rm FE} = 91$, the radial step $\Delta r = 1$ a.u. to the radial
boundary $r_{\rm max} = 91$ a.u. The total number of the radial basis
functions was $N_r = 364$.  An unphysical reflection from the upper
radial boundary was suppressed by the exterior complex scaling
(ECS). The ECS parameters were as follows: $r_{\rm ECS} = 51$~a.u. and
$\theta_{\rm ECS} = 30^\circ$. The angular variables were treated
separately with the number of azimuthal and polar points
$N_{\theta}=64$ and $N_{\phi}=128$, respectively. Unlike in the
previous work \cite{PhysRevA.84.062701}, we employ here the spherical
rather than spheroidal coordinates as a sufficiently large angular
basis was used to approximate the photoelectron wave packet far away
from the origin.  The ionization amplitude was extracted from the
time-dependent wave function using the time-corrected flux (t-SURFFc)
method) through a surface of a sufficiently large radius $r_S =
50$~a.u.

\begin{figure}[h]
\epsfxsize=6cm
\epsffile{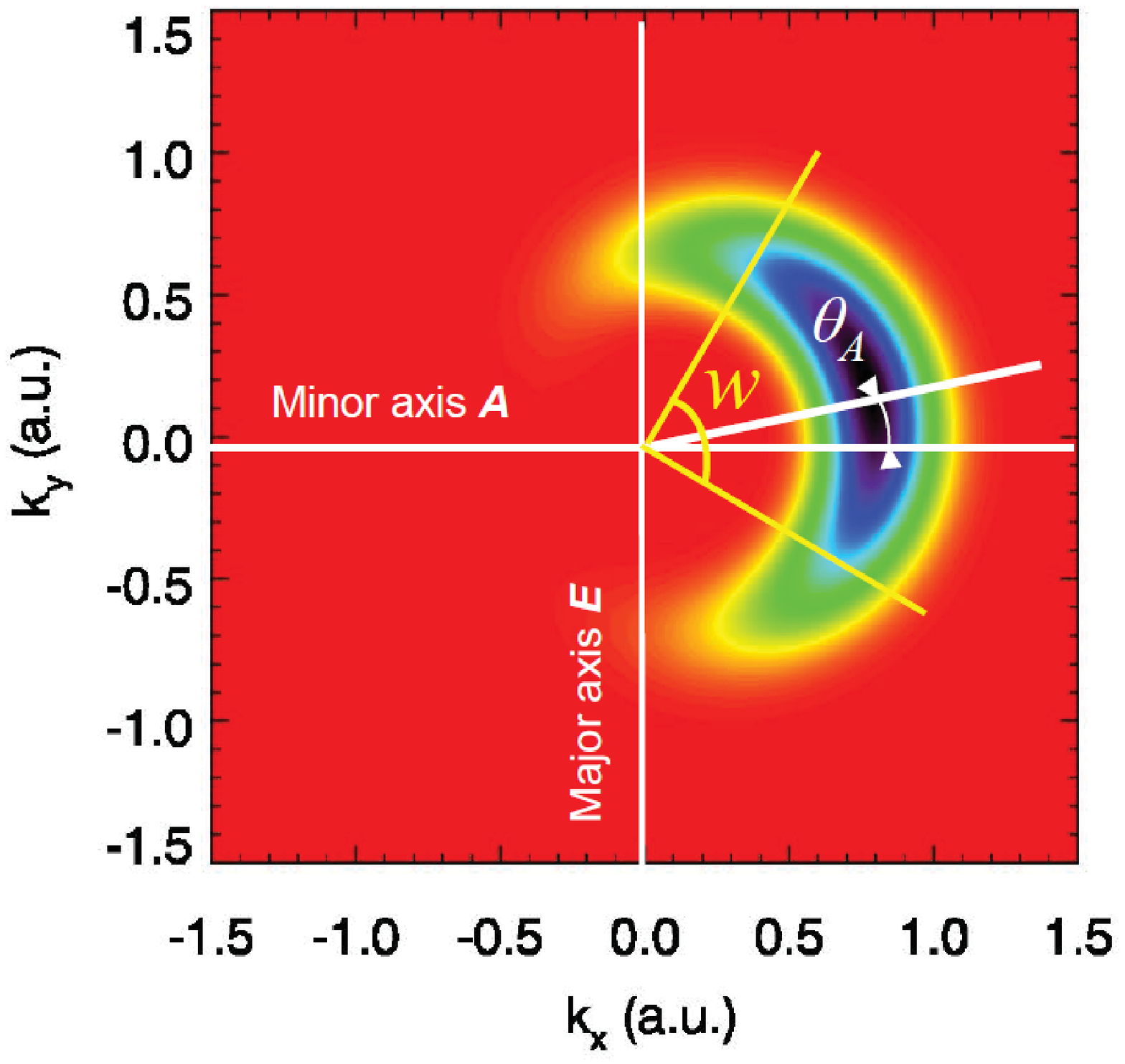}
\hs{.0cm}
\epsfxsize=6cm
\epsffile{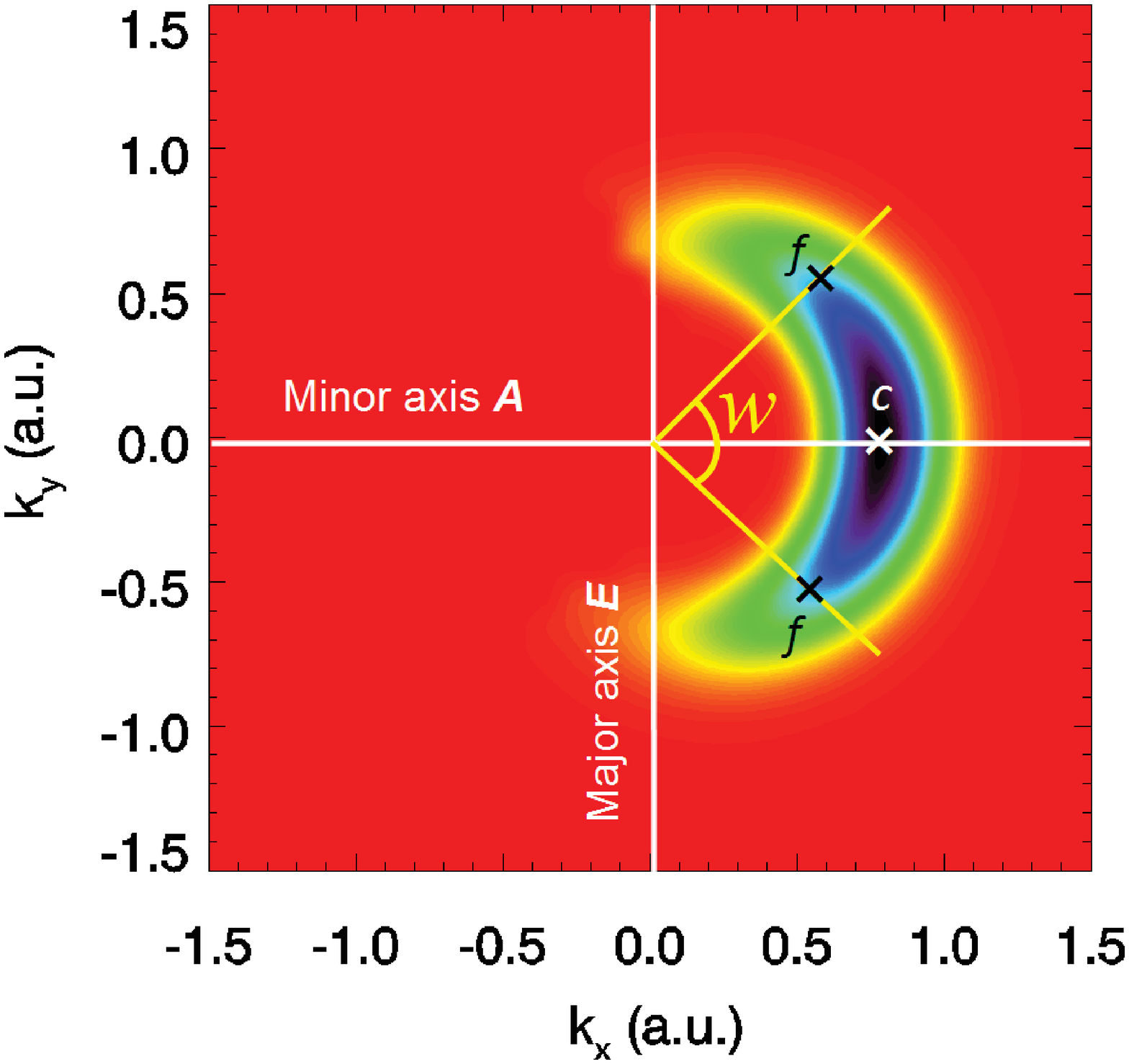}
\vs{-3mm}
\caption{Photoelectron momentum distribution $P(k_x,k_y,k_z=0)$ of the
  atomic hydrogen in the polarization plane at the driving field
  intensity $I=$\Wcm{8}{13}. The attoclock offset angle $\theta_A$ and
  the angular width $\cal W$ are marked. The coloration ranges from
  zero (red) to the maximum (black) linearly. Results of the
  TDSE (top) and SPM (bottom) calculations are displayed. The $c$ and
  $f$ labels on the bottom panel are used to mark the center and
  fringes of the PMD.
\label{Fig1}}
\end{figure}

A typical 2D momentum distribution in the polarization plane
$P(k_x,k_y,k_z=0)$ is shown for atomic hydrogen on the top panel of
\Fref{Fig1}.  This distribution is integrated radially to obtain the
angular distribution
$
P(\theta) = \int k\,dk \ P(k_x,k_y,k_z=0).
$
It is then fitted with a Gaussian 
$
P(\theta)\propto \exp[(\theta-\theta_A)^2/{\cal W}^2]
$
to determine the peak position and this value is
assigned to the attoclock offset angle $\theta_A$. The symmetry of
$P(\theta)$ relative to $\theta_A$ is carefully monitored and serves
as a test of the quality of the TDSE calculation. We also analyze the
Gaussian width $\cal W$ of this distribution. 

\subsection{Semi-classical approach}
\label{Classical}

We follow \cite{PhysRevLett.89.153001,0953-4075-39-14-R01}
and write the ionization amplitude as
\ba
\label{SPM}
D(\k) &=& -i\sum_{s=1}^{N_{\rm SP}}
\left\{
2\pi i\over \E(t_s)\cdot [\k+\A(t_s)]
\right\}^{1/2}
\\\nn &\times&
\la\k+\A(t_s)|\r\cdot\E(t_s)|\psi_0\ra \
\exp[iS_{\k}(t_s)] 
\ ,
\ea
where 
$
S_\k(t) = \int^t dt'
\{
[\k+A(t')]^2/2+I_p
\}
$
is the semiclassical action  
and, for hydrogenic targets, 
\be
\la\k|\r\cdot\E(t)|\psi_0\ra = -i 2^{7/2}(2I_p)^{5/4}
{\k\cdot \E(t)\over 
\pi(\k^2+2I_p)^3} \ .
\label{hydrogenic}
\ee
The summation in \Eref{SPM} is carried over the saddle
points $t_s$ that are solutions of the saddle point equation
\be
\label{SP}
\partial S_{\k}(t_s)/\partial t
=
[\k+\A(t_s)]^2/2+
I_p =0
\ .
\ee
Combining Eqs.~(\ref{SPM})-(\ref{hydrogenic}) under  condition
\eref{SP} leads to
\be
\label{SPM1}
\hs{-2mm}
D(\k) = -2^{-1/2}(2I_p)^{5/4}
\sum_{s=1}^{N_{\rm SP}}
\Big[S^{\prime\prime}_{\k}(t_s)\Big]^{-1}
\exp[iS_{\k}(t_s)]
\ .
\ee
The real part of the saddle points Re~$t_s$ specify those times when
the electron exits the tunnel to reach the detector with the drift
momentum $\k$. 
%
%
For circular polarization, the number of the saddle points $N_{\rm
  SP}=N+1$, where $N$ is the number of the pulse oscillations
\cite{0953-4075-39-14-R01}. With the presently chosen envelope,
$N_{\rm SP}=3$ of which only one dominant SP makes the overwhelming
contribution to the PMD shown on the bottom panel of \Fref{Fig1}.

\begin{figure}[t]
\epsfxsize=4.5cm 
\epsffile{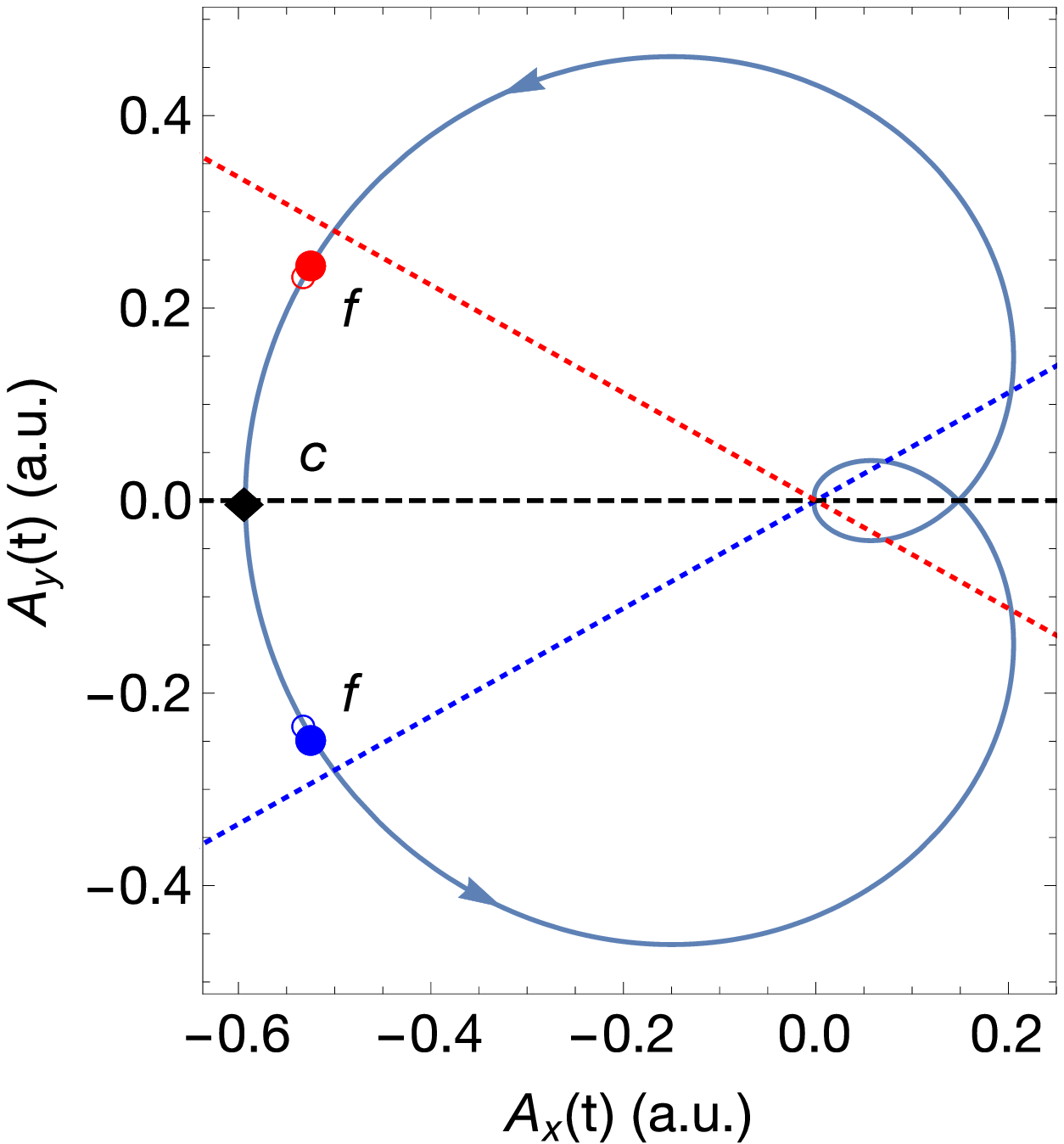}

\epsfxsize=4.5cm 
\epsffile{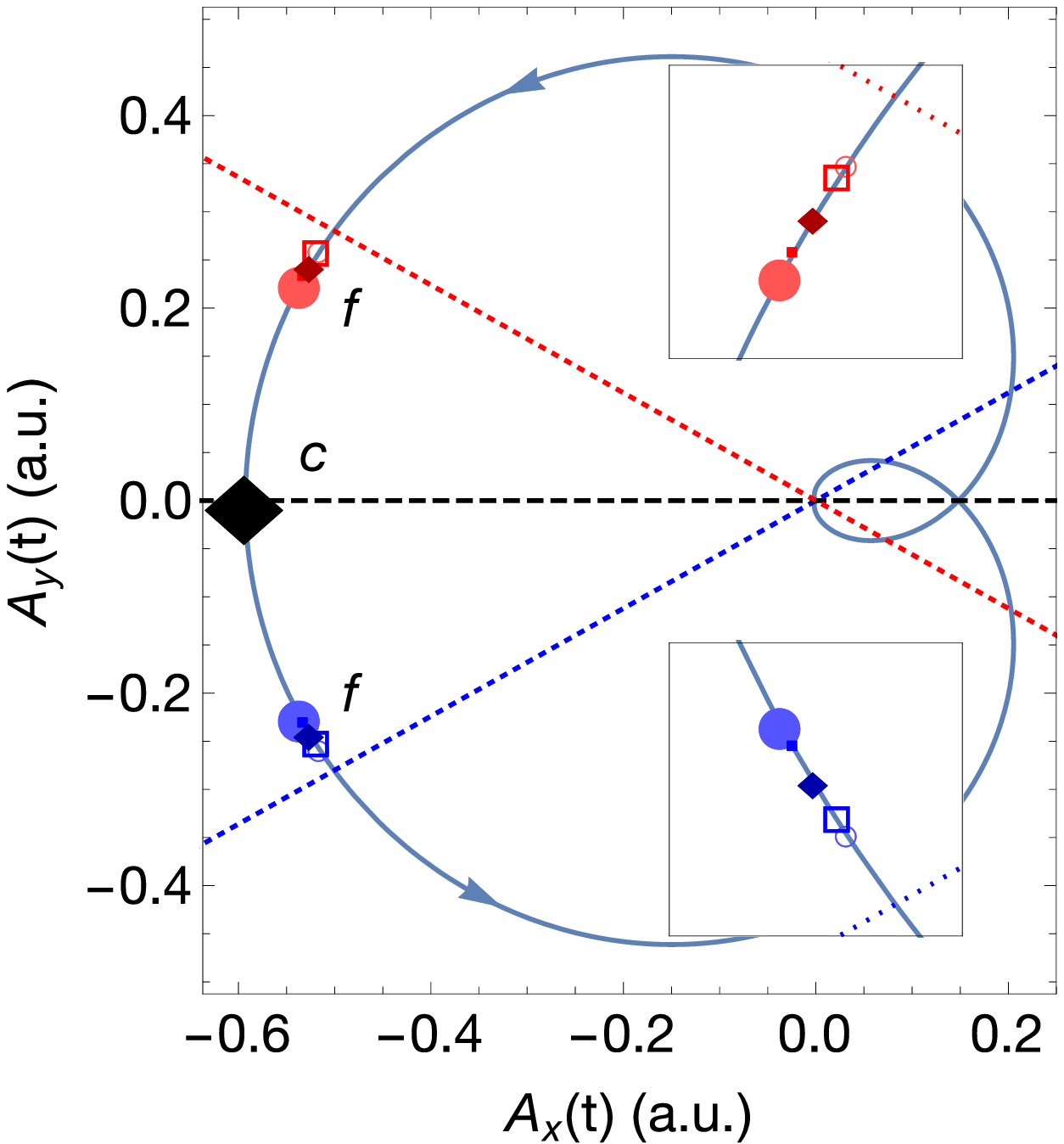}

\vs{-3mm}
\caption{ Saddle point solutions corresponding to the fringes $f$
  (blue/red) and center $c$ (black) of the PMD as marked on the bottom
  panel of \Fref{Fig1}.  The direction of each of the chosen momenta
  $\hat\k$ are given by dashed lines of the  matching
  color.  Top: atomic hydrogen (\ref{SP}) with $k_z=0.0$ a.u. (filled
  circles) and $k_z=1.0$ a.u. (open circles).  Bottom: molecular
  hydrogen with $j=1,2$ (\ref{SPH}) given by filled and open symbols,
  respectively.  The perpendicular orientation is given with diamonds
  of dark hue, the major axis orientation with circles of light hue,
  and the minor axis orientation with squares of standard hue.  The
  insets highlight the solutions about each fringe.  
\label{Fig2}}
\vs{-3mm}
\end{figure}

These saddle points are mapped on the parametric plot of the vector
potential $\A(t)$ in the polarization plane shown on the top panel of
\Fref{Fig2}.  Here the momenta $\hat\k$ corresponding to the center
$c$ and fringes $f$ of the PMD displayed on the bottom panel of
\Fref{Fig1} are drawn with straight lines.  In the absence of the
ionization potential, $\k=-\A(t_s)$, the saddle points are entirely
real and can be mapped on the intersection of the momentum and the
vector potential lines.  With a finite ionization potential, the
saddle points acquire an imaginary part and deviate from the
$\hat{\k}$ vector direction. This imaginary part Im~$t_s$ dampens the
contribution of a given saddle point exponentially.  In \Fref{Fig2} we
depict this relation via the point size which is proportional to 
$\exp(-{\rm Im}~t_s)$ normalised to that of the dominant, center saddle
point.  The minor saddle points near the origin have large Im~$t_s$
and hence their marks are not visible in the scale of the figure. The
relative size of the fringe saddle points affects the angular width of
the PMD.  The larger is Im~$t_s$ and the smaller the dots at the
fringes relative to that at the center, the smaller is the width.

In the case of the \H molecule, the SP equations \eref{SP} and
\eref{SPM1} require modification \cite{PhysRevA.73.023410}.  The
ground state wave function is expressed as a linear combination of the
hydrogenic orbitals
$
\psi_0 = C_\psi
\sum_{j=1,2}
\phi\left[\r+(-1)^j\R/2\right]
,
$
where $j$ is the atomic site index, $\R$ is the internuclear
separation vector, and $C_\psi$ is the overlap of the  orbitals
centered on the two atomic sites.
Accordingly, the ionization amplitude is written as
a coherent sum
\be
D(\k) = 
\sum_{j=1,2} D_j(\k,t_s) \exp\left[(-1)^j \ i\k\cdot\R/2)\right]
\ .
\label{HAMP}
\ee
Here $D_j(\k,t_s)$ denotes the ionization amplitude \eref{SPM1}
evaluated at the saddle point $t_s$ found as a solution of the 
modified SP equation 
\be
\big[
\k+\A(t_s)
\big]^2/2
+I_p - (-1)^j
\E(t_s)\cdot\R/2 = 0\ .
\label{SPH}
\ee
The difference between the atomic SP equation \eref{SP} and its
molecular counterpart \eref{SPH} is in the potential energy  term
$
\pm\E(t_s)\cdot\R/2 \ .
$
This term defines the energy gain or loss for the electron to travel
to the molecule mid point and has the opposite signs for different
atomic sites. Accordingly, the single dominant SP in the atomic case
is split into two points as illustrated graphically on the right panel
of \Fref{Fig2}. The corresponding factor 
$
\exp[\pm i \k\cdot\R/2]
$
in the ionization amplitude \eref{HAMP} defines the phase difference
between the two wave packets emitted from different atomic sites. 
We note that the molecular terms in both Eqs.~\eref{HAMP} and \eref{SPH}
vanish when the molecule is aligned perpendicular to the polarization
plane.

\section{Results}
\label{Results}

\subsection{Atomic hydrogen}
\label{Atomic}

\begin{figure}[h]
\hs{-5mm}
\epsfxsize=9.5cm 
\epsffile{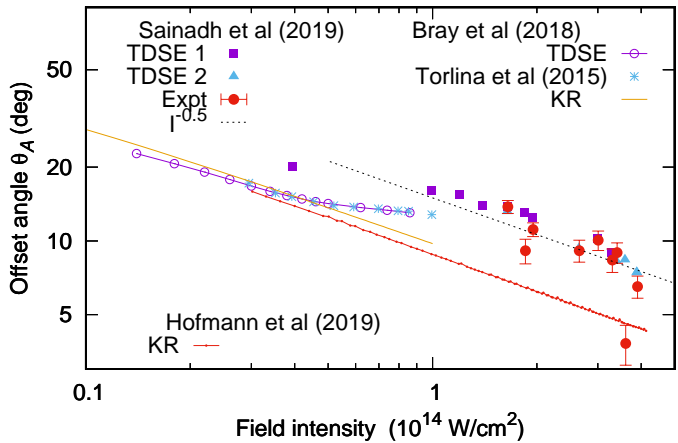}

\caption{The attoclock offset angle $\theta_A$ plotted as a function
  of the laser field intensity. The TDSE1 (red squares) and TDSE2
  (blue triangles) calculations and the experimental points (with
  error bars) visualize the data of \citet{Sainadh2019} obtained with
  multi-cycle pulses. The TDSE calculation (red open circles) and the
  KR model prediction (red solid line) from
  \citet{PhysRevLett.121.123201} refer to the single-oscillation
  pulses. Similar calculation of \citet{Torlina2015} (blue asterisks)
  is also shown.  The KR model is extended beyond the range of its
  validity by \citet{Hofmann2019} and displayed with the thick dotted
  line. The thin dotted line fits the data of \citet{Sainadh2019} with
  the characteristic field intensity dependence $I^{-0.5}$ predicted
  by the KR model.
\label{Fig0}}
\end{figure}

The attoclock offset angles $\theta_A$ as functions of the field
intensity under different driving pulse conditions are exhibited in
\Fref{Fig0}. Here we display the experimental and theoretical results
of \citet{Sainadh2019} obtained with multi-cycle CEP averaged
pulses. The presently employed iSURF code was calibrated against the
TDSE1 and TDSE2 calculations of \citet{Sainadh2019} at several field
intensity and CEP values. The corresponding results were found to be
fully compatible.  In the same figure, we present the TDSE
calculations with single-oscillation pulses
\cite{PhysRevLett.121.123201, Torlina2015} which are very similar to
each other but deviate noticeably from the TDSE1 and TDSE2 results
corresponding to CEP averaged multi-cycle elliptical pulses. In the
low field intensity range, the TDSE results merge the predictions of
the KR model \cite{PhysRevLett.121.123201} which was designed to
explain a very steep rise of the offset angle $\theta_A$ with
decreasing laser pulse intensity. It was tempting to explain this rise
in terms of the increasing tunnel with and corresponding increase of
the tunneling time. However, in reality, this is an effect of the
stronger elastic scattering of a slower photoelectron in the Coulomb
field of the residual ion. We note that the KR model deviates
noticeably from the TDSE at an increasing laser field. It is therefore
meaningless to extend it to very large intensities as was done in
\cite{Hofmann2019}. Moreover, it is even less meaningful to explain
the difference between the KR predictions and the experiment by a
finite tunneling time as was suggested in \cite{Hofmann2019}. The
offset angles became virtually zero when the long-range Coulomb field
was replaced by a short-range Yukawa potential in both TDSE1 and TDSE2
calculations \cite{Sainadh2019}.

\begin{figure}[h]
\epsfxsize=7cm 
\epsffile{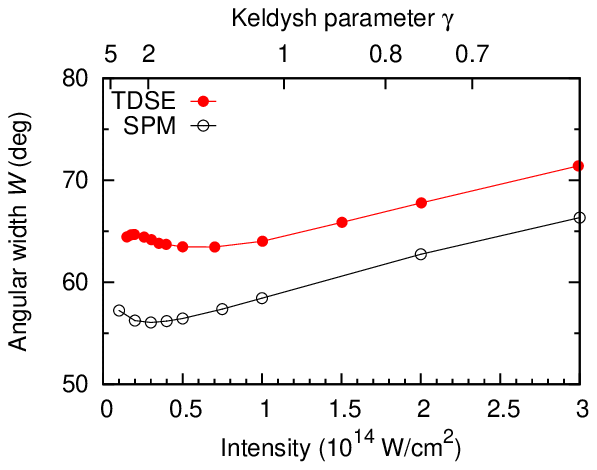}
\epsfxsize=7cm 
\epsffile{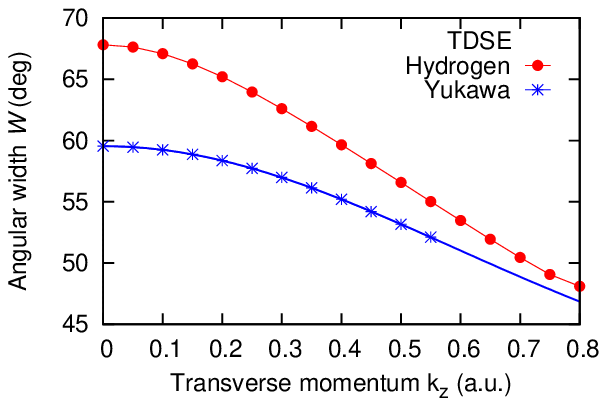}
\vs{-3mm}
\caption{Top: Angular width ${\cal W}(I)$ of the PMD of atomic
  hydrogen in the polarization plane as a function of the field
  intensity $I$. Bottom: The width parameter ${\cal W}(k_z)$ as a
  function of the transverse photoelectron momentum outside the
  polarization plane at the field intensity of \Wcm{2}{14}.
\label{Fig3}}
\end{figure}

The main difference between the PMD shown on the top and bottom
panels of \Fref{Fig1} is that $\theta_A=0$ in the SFA and SPM.  This
is another indication that the main contribution to the attoclock
offset angle comes from the Coulomb field of the residual ion which is
not included in the SFA. Except for this offset angle, the overall
structure of the PMD in the polarization plane is reproduced
remarkably well by the SPM. We quantify this PMD by its angular
width $\cal W$ which we extract from the Gaussian fitting to the
radially integrated momentum density. The width parameter extracted
from the TDSE and SPM calculations are shown in \Fref{Fig3}. On the
top panel, we show the width ${\cal W}(I)$ in the polarization plane
as a function of the field intensity $I$. We observe that this
dependence is not monotonous. This feature can be qualitatively
understood from the SFA formulas given in \cite{Mur2001} for a
continuous elliptical field. In this case, the SP equation \eref{SP}
can be solved analytically.  For strong fields, when the Keldysh
adiabaticity parameter $\gamma\ll1$, the angular width grows with
intensity as ${\cal W}\propto 1/\sqrt{\gamma}\propto I^{1/4}$. In the
opposite limit $\gamma\gg1$ the width is falling with intensity as
${\cal W}\propto \sqrt{\ln(c\gamma)}$ with $c=2/(1-\xi^2)^{1/2}$
expressed via the ellipticity parameter $\xi$. The cross-over between
these falling and rising intensity dependence of the width occurs
around $\gamma\simeq 1$.

On the bottom panel of \Fref{Fig3} we display the width dependence on
the transverse momentum outside the polarization plane ${\cal W}(k_z)$
at a fixed field intensity. We see that the width is rapidly falling
with increasing $k_z$. A significant part of this fall is reproduced
by the SPM calculation simply because the SP equation \eref{SP}, which
can be understood as the energy conservation at the exit from the
tunnel, contains an additional kinetic energy term $k_z^2/2$. This
increases an effective ionization potential
\cite{PhysRevLett.121.163202} and thus reduces Im~$t_s$ as illustrated
on the top diagram of \Fref{Fig2} (filled symbols at $k_z=0$ versus
open symbols at $k_z=1$~au). Hence the angular width decreases.  The
effect of the Coulomb field of the proton is seen clearly from
comparison of the two TDSE calculations on the hydrogen and Yukawa
atoms. In the latter case, the Coulomb field is screened
$V(r)=-(Z^*/r)\exp(-r/2)$ with $Z^*=1.47$ to maintain identical
ionization potentials. The TDSE calculation on the Yukawa atom is
particularly close to the SPM result. Such a calculation also exhibits
a zero angular offset $\theta_A=0$
\cite{Torlina2015,PhysRevLett.121.123201,Sainadh2019}.

\subsection{Molecular hydrogen}
\label{Molecular}

\begin{figure}[h]
\vs{4cm}

\epsfxsize=7cm 
\epsffile{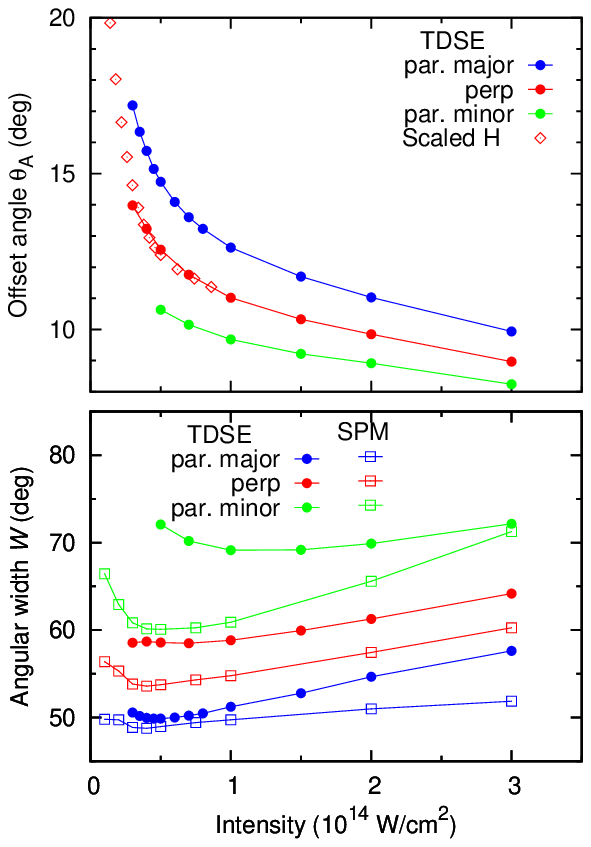}
\vs{-3mm}

\caption{The attoclock offset angle $\theta_A$ (top) and the angular
  width $\cal W$ (bottom) of the \H molecule as functions of the field
  intensity. The molecule is aligned perpendicular to the polarization
  plane (red), parallel to the major polarization axis $\hat{\bm{e}}_y$
  (black) and parallel to the minor polarization axis $\hat{\bm{e}}_x$
  (green). The top panel shows the offset angle $\theta_A$ of the H
  atom scaled to the ionization potential of \H.
\label{Fig4}}
\end{figure}

We illustrate the effect of the molecular orientation in \Fref{Fig4}
where we show the attoclock offset angle $\theta_A$ (top) and the
angular width parameter $\cal W$ (bottom) for the \H molecule in three
orientations: perpendicular to the polarization plane (shown with red
symbols), aligned with the peak $\E$ field (``major'' $\hat{\bm{e}}_y$
axis, blue symbols) and with the peak $\A$ potential (``minor''
$\hat{\bm{e}}_x$ axis, green symbols).  Both the TDSE and SPM results
are shown for the width parameter (filled circles and open squares,
respectively).
We observe on the top panel that the attoclock offset angles
$\theta_A$ vary minorly with the molecular orientation. On the
same panel we plot the corresponding offset angles of the atomic
hydrogen. According to the KR model, the offset angle due to the
scattering in the Coulomb potential is estimated as
%
$
\theta_A=
(\w^2/ E_0)(Z/ I_p) \ ,
$
%
where $Z$ is an effective charge of the residual ion. To bring the
offset angle of the hydrogen atom to the \H scale, we need to multiply
it by the corresponding ionization potentials ratio 
$
\theta_A^{\rm H} 
\times
I_p^{{\rm H}}
/I_p^{{\rm H}_2}
,
$
where for the \H molecule we take the HF ionization potential
$I_p^{{\rm H}_2}=15.6$~eV. We see that such a scaled
atomic hydrogen calculation is almost indistinguishable from the \H
results in the perpendicular orientation for which the interference
terms 
%
%
vanish. 

The angular width $\cal W$ varies very significantly depending on the
molecular orientation. This behaviour is similar in the TDSE and SPM
calculations.  The latter model allows for the understanding of this
behavior qualitatively in terms of the two-center interference through
analysis of the underlying saddle points, illustrated in the right
panel of \Fref{Fig4}.  For a given in-plane orientation the
interference term in \Eref{SPH} effectively increases (decreases) the
ionization potential thus increasing (or decreasing) the Im~$t_s$ and
relative contribution of the corresponding saddle points.

\begin{figure}[t]

\vs{3.5cm}
\epsfxsize=7.cm 
\epsffile{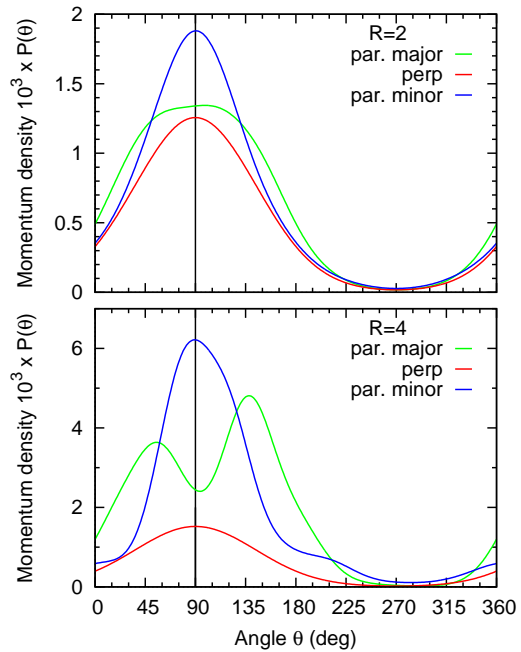}

\caption{Radially integrated PMD $P(\theta)$ in the polarization plane
for the three orientations of the Yukawa molecule with
$R=2$~a.u. (top)  and 4~a.u. (bottom).
\vs{-2mm}
\label{Fig5}}
\end{figure}

More insight into the orientation dependence of the attoclock width
$\cal W$ can be gained from a model Yukawa molecule in which the two
Yukawa atoms are placed at a varying distance. Results of the TDSE
calculation for such a target are exhibited in \Fref{Fig5}. Here we
display the angular distribution $P(\theta)$ of the radially
integrated PMD in the polarization plane for alternate molecular
orientations and the two inter-atomic distances of $R=2$ and
4~a.u. For the ``major'' axis molecular alignment, the potential energy
term in the SP equation \eref{SPH} is large. It suppresses
photoemission from one atomic site and enhances it from another. The
enhanced photoelectron wave packet has a smaller angular width which
is carried through to the detector. For the ``minor'' axis alignment, the
two photoelectron wave packets have the same width but are displaced
relative to the detection direction of $90^\circ$ to the opposite
sides. This results in a visibly non-Gaussian shape and a considerable
increase of the width of the PMD at a small separation
$R=2$~a.u.\ evolving into two split  peaks at a larger separation $R=4$~a.u. 
A non-gaussian shape of the PMD may explain why the agreement of the
TDCS and SPM calculations for the ``major'' and ``minor'' axis
alignment on the bottom panel of \Fref{Fig4} is poorer than for
perpendicular molecular orientation.
Lack of symmetry of the PMD peak relative to the
$90^\circ$ direction may also explain a small deviation of the
attoclock offset angles at different molecular orientations visible on
the top panel of \Fref{Fig4}.

\subsection{Neon dimer}
\label{Dimer}

The hint of the two-center Young-type interference in \H, which
manifests itself as the orientation dependent angular width $\cal W$,
is greatly exemplified in a model Yukawa molecule with an enlarged
inter-atomic distance. This enlargement happens naturally in weakly
bound noble gas dimers held by weak van der Waals
forces. Consequently, the photoelectron momentum distribution shows a
very distinct interference pattern. This pattern depends strongly on
the symmetry of the molecular orbital being ionized which is captured
by the kinetic energy release (KER) from the corresponding ionic
dissociative state
\cite{Kunitski2019}.

In these measurements, performed on the \N dimer with a long, 40~fs
circularly polarized laser pulse, the KER of the
dissociating atomic fragments were analyzed. Thus, the gerade
$2p\sigma_g$ and ungerade $2p\sigma_u$ orbitals of the molecular
ground state could be readily resolved by their distinctively 
different KER. It
was argued that the corresponding two-center interference factor
\be
\label{interference}
P\propto  \cos^2 
\left(
\k\cdot {\R\over 2}
+
{\Delta\phi\over 2}
\right)
\ee
should contain the phase difference $\Delta\phi=\pi$ (gerade) and 
$0$ (ungerade), accordingly. 
When the two symmetries are mixed, the fringes and
anti-fringes overlap and the interference pattern is washed away.

\begin{figure}[t]

\epsfxsize=6.cm 
\epsffile{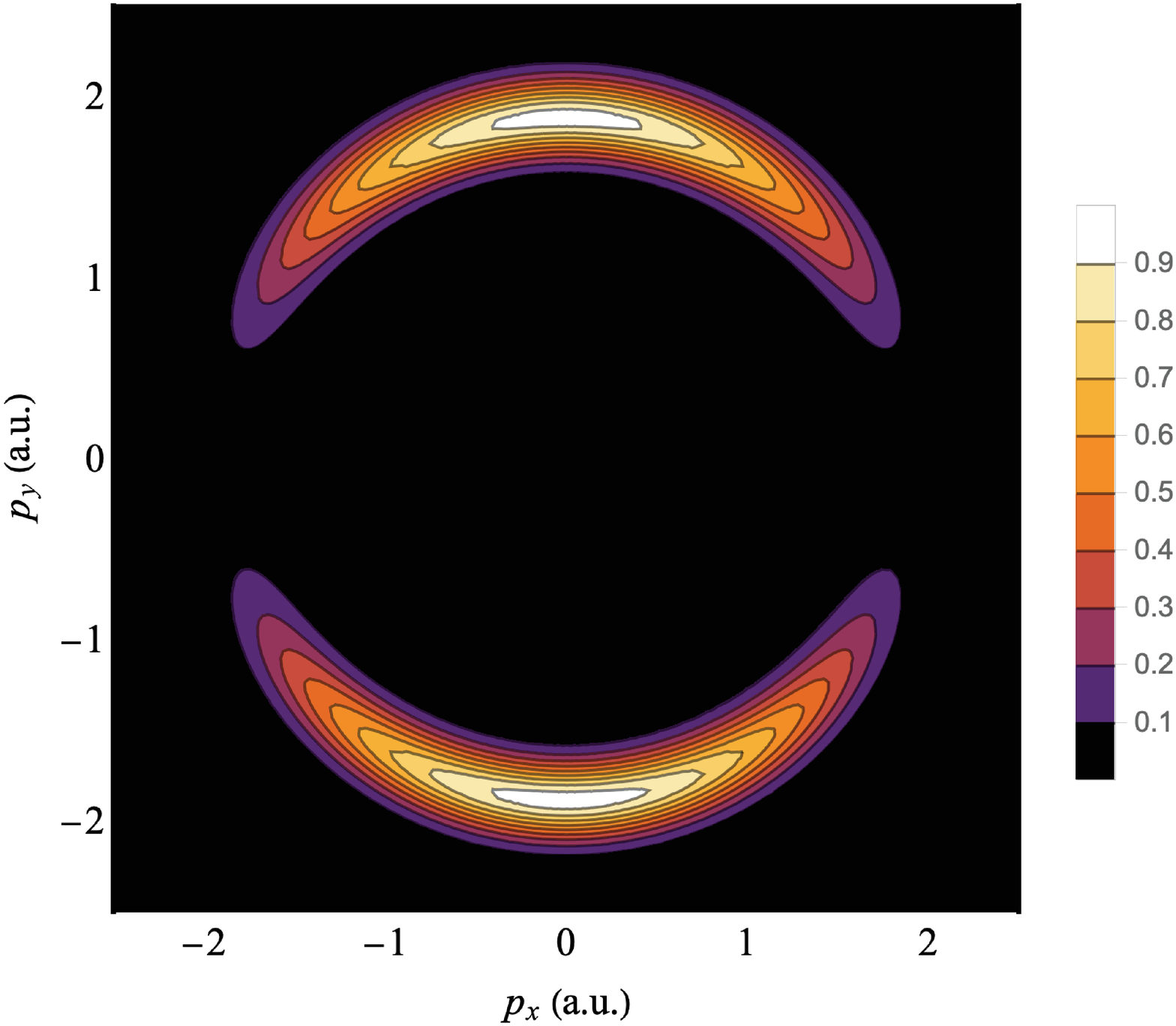}

\epsfxsize=6.cm 
\epsffile{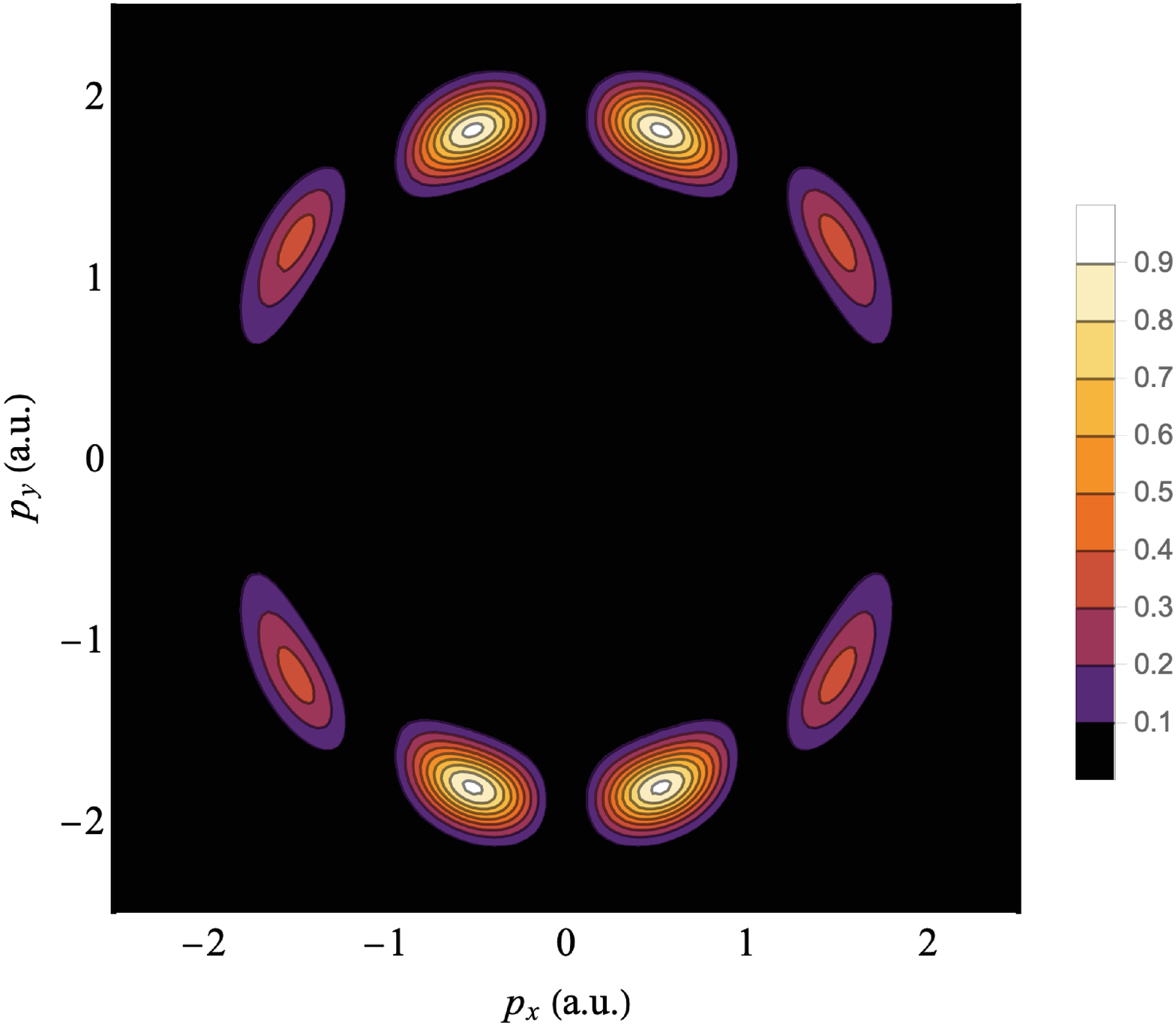}

\epsfxsize=6.cm 
\epsffile{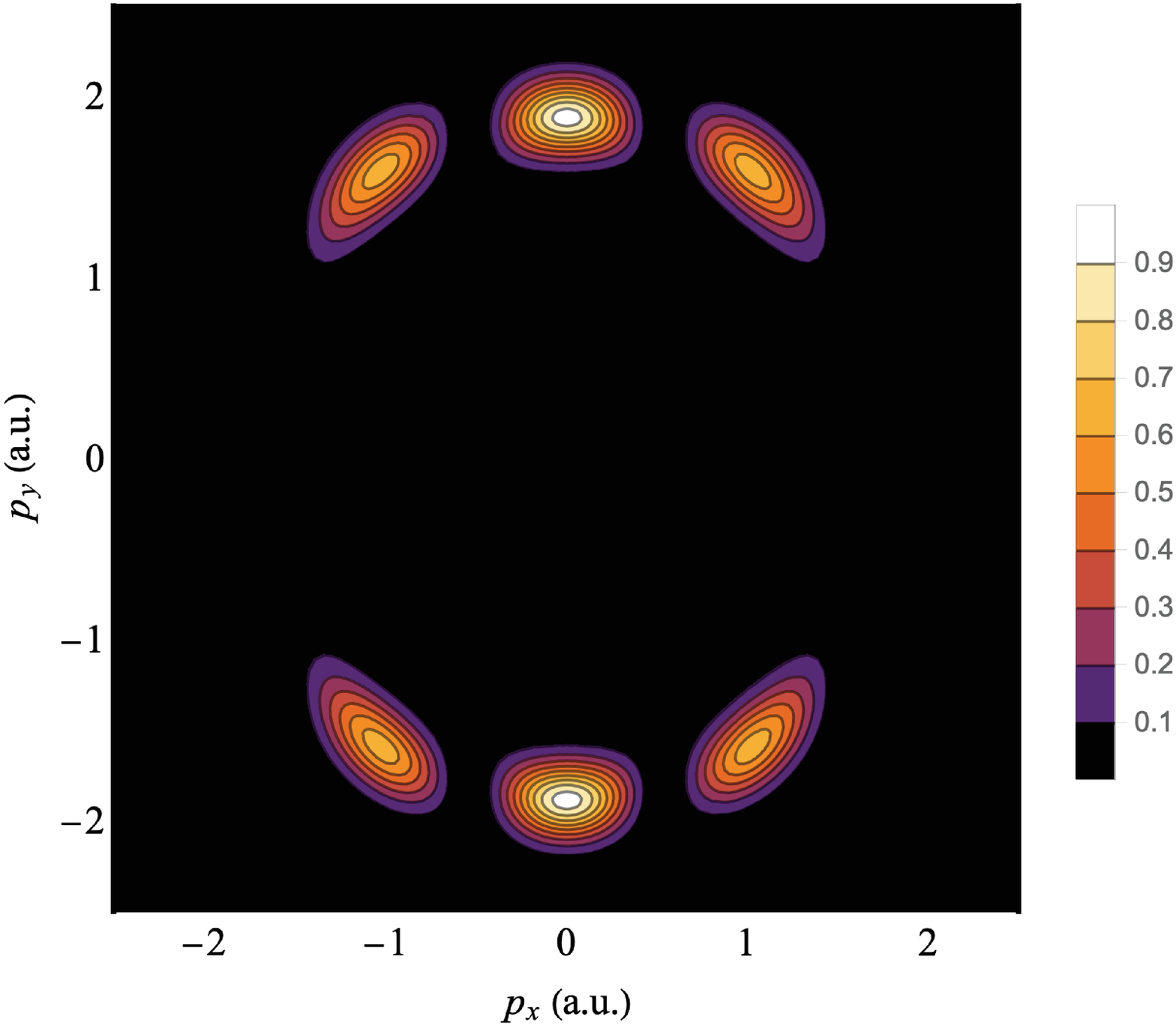}

\caption{Photoelectron momentum distribution in \N in the polarization plane
of the circularly polarized continuous laser field.  Top: atomic-like SPM
calculation using analytic expressions of \cite{Mur2001}. Middle and bottom: the
same calculation modulated by the interference factor
\eref{interference} with the phase difference $\Delta\phi=\pi$ (gerade) and 
$0$ (ungerade), respectively.
\vs{-2mm}
\label{Fig6}}
\end{figure}

This behavior is seen on the top panel of \Fref{Fig6} where we
generate an atomic-like momentum distribution in the polarization
plane of a continuous circularly polarized field.  Here we take the
ionization potential of the Ne atom $I_p=21.6$~eV, $\omega=0.0584$
a.u.\ ($\lambda=780$~nm), peak intensity $7.3\times10^{14}$~W/cm$^2$, and employ
the analytical SPM formula given by Eq.\ (27) of \cite{Mur2001} multiplied 
by a spherical harmonic.

We then take the atomic ionization amplitude and insert it into the molecular
expression \eref{HAMP} with $\bm{R}=5.85\hat{e}_x$ a.u.\ and get the
momentum distribution shown in the middle panel of \Fref{Fig6}. If we
drop the $(-1)^j$ factor in \Eref{HAMP}, we generate the momentum
distribution shown on the bottom pane. The keeping and dropping of the
phase factor $(-1)^j$ in \Eref{HAMP} is equivalent to selecting the
phase difference $\Delta\phi=\pi$ and $0$, accordingly. We find 
these three distinctively different distributions to 
correspond very closely with the experimental plots shown in the inset
of Fig.~1, and Figs.~2(b) and 2(e) of \cite{Kunitski2019},
respectively.

\section{Conclusion}
\label{Conclusion}

In conclusion, we conducted a numerical study of attoclock
on atomic and molecular hydrogen driven by a single-cycle circularly
polarized laser pulse. We interpreted our numerical results using the
simplified semiclassical SPM and the classical KR models. The latter
model treats the attoclock as a ``nano-ruler'' and relates the offset
angle of the peak photoelectron momentum distribution with the tunnel
width rather than the tunneling time. We explore the range of validity
of the KR model and compared theoretical and experimental results
obtained with short, nearly single-oscillation, and longer
multi-oscillation pulses.  By analyzing the width of the photoelectron
momentum distribution in the \H attoclock, we found a clear signature
of the two-center electron interference. This celebrated effect has
many decades of history but it is the present work that revealed it
for the molecular attoclock.  The interference effect is encoded in
the molecular saddle point equation \eref{SPH} and reflects the
bonding or anti-bonding nature of the molecular ground state.  We
demonstrated this effect in the \H molecule, an artificially enlarged
Yukawa molecule and the naturally extended N$_2$ dimer which has a
very large inter-atomic separation.

The authors are greatly indebted to Serguei Patchkovskii who placed
his iSURF TDSE code at their disposal.

The authors wish to thank Satya Sainadh, Igor Litvinyuk, Robert Sang
and Sebastian Eckart for many stimulating discussions.  Resources of
the National Computational Infrastructure were employed.



\end{document}